\documentclass[a4paper]{article}

\usepackage{amsmath}
\usepackage{amsthm}
\usepackage{amssymb}
\usepackage{authblk}

\usepackage{bm}

\usepackage[utf8]{inputenc}
\usepackage{hyperref}
\hypersetup{
	unicode,
	pdfauthor={Author One, Author Two, Author Three},
	pdftitle={A simple article template},
	pdfsubject={A simple article template},
	pdfkeywords={article, template, simple},
	pdfproducer={LaTeX},
	pdfcreator={pdflatex}
}

\usepackage[sort&compress,numbers,square]{natbib}
\theoremstyle{plain}
\newtheorem{theorem}{Theorem}

\newtheorem{lemma}[theorem]{Lemma}

\newtheorem{proposition}[theorem]{Proposition}

\theoremstyle{definition}
\newtheorem{definition}[theorem]{Definition}

\newtheorem{remark}[theorem]{Remark}

\usepackage{graphicx, color}
\graphicspath{{fig/}}

\usepackage{algorithm, algpseudocode} % use algorithm and algorithmicx for typesetting algorithms
\usepackage{mathrsfs} % for \mathscr command

\usepackage{lipsum}

\title{Abelian and consta-Abelian polyadic codes over affine algebras  with a finite commutative chain coefficient ring
}
\author[a]{Gülsüm Gözde Yılmazgüç \thanks{This author is supported by TÜBİTAK within the scope of 2219 International Post Doctoral Research Fellowship Program with application number 1059B192101164. Her work was completed  while she visited the Institute of Mathematics of University of Valladolid (IMUVa). She thanks the IMUVa for their kind hospitality.}}
\author[b]{Javier de la Cruz} 
\author[c]{Edgar Mart\'inez-Moro\thanks{Partially supported by Grant TED2021-130358B-I00 funded by MCIN/AEI/10.13039/501100011033 and by the “European Union NextGenerationEU/PRTR”}}

\affil[a]{Ipsala Vocational College, Trakya University, Edirne, Turkey}
\affil[b]{Departamento de Matem\'aticas, Universidad del Norte,   Barranquilla, Colombia}
\affil[c]{Institute of Mathematics, University of Valladolid, Castilla, Spain }

\date{\today
}

\begin{document}
	\maketitle
	
	\begin{abstract}
  In this paper,  we define  Abelian and consta-Abelian polyadic codes over rings defined as affine algebras over chain rings. For that aim, we use the classical construction  via splittings and multipliers of the underlying Abelian group.  We also derive some results on the structure of the associated polyadic codes and the number of codes under these conditions.\vspace{0.5cm}	
  
	\end{abstract}

%	\tableofcontents
	
	\section{Introduction}
	\label{sec:intro}
	Polyadic codes  were first introduced in \cite{Brualdi}.
	There is a rich literature on this type of codes, see for example \cite{Chen,Lim,Ling1} and the references therein. Recently in \cite{Indian,Indian2} they extended some of these ideas to codes over rings.
	
	In this paper we extend those results in a twofold way: we answer some questions settled in \cite{Ling1}, namely the generalization of polyadic  Abelian codes to the case of chain rings 
and, also to consider the more general case of considering a class of serial rings as ambient space that do not entirely split into linear factors over the base chain ring. Up to our knowledge, this type of base rings have been considered only for the case that the polynomials defining it split completely into linear factors and  the  underlying ring is a finite field, see \cite{Indian, Indian2}.  Note that general linear codes over this type of rings, namely affine algebras with a finite commutative chain
              coefficient ring  was already studied in \cite{algebras} and a concrete example of those ambient spaces was studied in \cite{Habibul} that is closely related to the construction in \cite{Indian, Indian2}, but in \cite{algebras} only general linear codes are studied and there is no advantage taken of the underlying group structure as in the present paper.
	
	 The outline of the paper will be as follows. In Section~\ref{sec:pre}, we  introduce some preliminaries on finite chain rings and serial polynomial rings over them as well as their idempotents. Section~\ref{sec:codesovermathcal} is devoted to codes over those types of rings, we will take particular care of the structure of constacyclic codes and multiconstacyclic codes. In Section~\ref{sec:polychain}, we defined Abelian and consta-Abelian polyadic codes over chain rings via splittings and multipliers. Sections~\ref{sec:core1}~and~\ref{sec:core2} are the core part of the paper where we study Abelian and consta-Abelian polyadic codes over affine alebra rings  with a finite commutative chain
coefficient ring. We finish   with some conclusions in Section~\ref{sec:conclude}.
  
	\section{Preliminaries}
	\label{sec:pre}
	
	In this section, we will fix our notation and recall some basic facts about finite chain rings (see for example \cite{McDonald} for a complete account) and serial polynomial rings over a chain ring (see \cite{Serial}). In this paper, all rings will be associative, commutative, and with identity. A ring $R$ is called a local ring if it has a unique maximal ideal. A local ring is a chain ring if its lattice of ideals is a chain under inclusion. In this case, since the ideals are linearly ordered by inclusion, the ring is also called uniserial.  It can be shown [8, Proposition 2.1] that $R$ is a finite commutative chain ring if and only if $R$ is a local ring and its maximal ideal is principal. We will denote by $a\in R$  a fixed generator of the maximal ideal $\mathfrak m$, and let $t$ be its nilpotency index, thus the ideals of $R$ are $\mathfrak m^i=\langle a^i\rangle$ for $i=0,\ldots,t$. Also, we will denote the residue field of $R$ by $\mathbb F_q= R/\mathfrak m$, where $q=p^l$, for a prime number $p$.  We will denote the polynomial ring in the indeterminates $X_i$ by $R[X_1,\ldots , X_s]$ for $i=1,\ldots s$ with its coefficients in $R$.  We can extend the natural ring homomorphism $ \bar .$ from $R$ to $\mathbb F_q$ given by $r\mapsto \bar r=r+\mathfrak m$ to the polynomial rings $R[X_1,\ldots , X_s]$ and $\mathbb F_q [X_1,\ldots , X_s]$ just by applying $\bar .$ on each coefficient of the polynomial. Let $t_i(X_i)\in R[X_i]$ ($i=1,\ldots,s$) be monic polynomials such that each  $\bar t_i(X_i)\in \mathbb F_q[X_i]$ is a square-free polynomial. During this paper, we are interested in codes over the following alphabet
	\begin{equation}\label{eq:curlyR}
	    \mathcal R= R[X_1,\ldots , X_s]/I \hbox{ where } I=\langle t_1(X_1),\ldots ,t_s(X_s) \rangle.
	\end{equation}
 Note that this setting includes as particular cases the alphabets considered in \cite{Indian, Indian2, Habibul}.
 In \cite{Serial}  the ideals of  $\mathcal R$ have been described explicitly.  This structure has been also studied in [18, 19], in the case where the ring $R$ is the finite field $\mathbb F_q$. In the finite field case, the square-free condition on the polynomials $t_i(X_i)$ is known as the ``semisimple condition''  because of the   structure of the ring $\mathcal R$ (it can be decomposed as a direct sum of simple ideals). In the general case, the square-free condition on the polynomials $t_i(X_i)$  leads to a decomposition of the ring $\mathcal R$ as a direct sum of finite chain rings, and therefore it is a serial ring \cite{SerialRing}. In the remaining part of the preliminaries, we will follow \cite{Serial} to explicitly show decomposition in terms of primitive idempotents. 
 
 Let $H_i$ be the set of roots of $\bar t_i(X_i)$ in a suitable extension of $\mathbb F_q$ for $i=1,\ldots, s$. 
 For each $\nu\in \mathcal H=\prod_{i=1}^s H_i$ we define the class of $\nu$ as
 $C(\nu)=\{(\nu_1^{q^j},\ldots , \nu_s^{q^j})\mid j\in \mathbb N\}$. We will denote the set of all the classes as $\mathcal C=\mathcal C(t_1,\ldots , t_s)$, the elements of $\mathcal C$ form a partition of $\mathcal H$ and for any ideal $I \triangleleft \mathcal R/\mathfrak m$ the set of the common zeros of the elements in $I$ is a union of classes. Also the size of each class is given by $|C(\nu)|=\mathrm{l.c.m.}(d_1,\ldots , d_s)=[\mathbb F_q(\nu_1,\ldots ,\nu_s):\mathbb F_q] $ where $d_i$ is the degree of the irreducible polynomial of $\nu_i$ over $\mathbb F_q$.
 
 For all $i=1,\ldots ,s$ and a class $C$, let $p_{C,i}(X_i)$ denote the polynomial $\mathrm{Irr}(\nu_i,\mathbb F_q)$ and $(\nu_1,\ldots ,\nu_s) \in C$ . Also  for all $i=2,\ldots ,s$ we consider the polynomials $b_{C,i}(X_i) = \mathrm{Irr}(\nu_i,\mathbb F_q(\nu_1,\ldots,\nu_{i-1}))\in \mathbb F_q(\nu_1,\ldots,\nu_{i-1})[X_i]$ and $\Tilde{b}_{C,i}(X_i)=\frac{p_{C,i}(Xi)}{b_{C,i}(Xi)}$. Note that it is clear that they are independent of the element $\nu$ chosen within $C$ and that ${b}_{C,i}(X_i)$ and $\Tilde{b}_{C,i}(X_i)$  are coprime polynomials. Then, define the multivariable  polynomials $w _{C,i} (X_1 , \ldots , X_i )$, and $\pi_{C,i}(X_1 , \ldots , X_i )$ obtained from ${b}_{C,i}(X_i)$ and $\Tilde{b}_{C,i}(X_i)$ respectively by substituting $\nu_i$ by $X_i$. One has that 
 $$\mathbb F_q [X_1 , \ldots , X_n ]/ \langle p_{C,1}, w_{C,2},\ldots , w_{C,n}\rangle\simeq
\mathbb F_q (\nu_1, \ldots ,\nu_n),$$ and we denote the Hensel lifts to $R$ of the polynomials $p_{C,i}$, $w_{C,i}$ and $\pi_{C,i}$ by $q_{C,i}$, $z_{C,i}$ and $\sigma_{C,i}$   respectively. If we denote by $I_C=\langle q_{C,1}, z_{C,2},\ldots , z_{C,s}\rangle$  then the ring $T_C = R [X_1 , \ldots , X_s ]/ I_C $ is a chain ring with maximal ideal $\mathfrak M =\langle a, q_{C,1}, z_{C,2},\ldots , z_{C,s}\rangle + I_C$ and 
$T_C /\mathfrak M\simeq   \mathbb F_q (\nu_1, \ldots ,\nu_s)$ (see \cite[Remark 4, \& Lemma 3.5]{Serial}). Now consider the polynomial 
$$ h_{C}(X_1,X_2,...,X_s)=\prod_{i=1}^s\frac{t_i(X_i)}{q_{\mathcal C , i}(X_i)}\prod_{i=1}^s\sigma_{q_{\mathcal C , i}(X_2,\ldots , X_i)}, $$ 
then $I_C+I=\mathrm{Ann}(\langle h_C+I\rangle)$, $\langle h_C+I\rangle\simeq R [X_1 , \ldots , X_s ]/ I_C$ and 
$\mathcal R\simeq \bigoplus_{C\in\mathcal C} \langle h_C+I\rangle$ (
for a proof see \cite[Proposition 3.7, Lemma 3.8 \& Theorem 3.9]{Serial}).
This
decomposition of the ring $\mathcal R$ is equivalent to the existence of primitive
orthogonal idempotents
elements $e_C \in \mathcal  R$ where $C\in \mathcal C$  such that $1 =\sum_{C\in \mathcal C}e_C$ 
and $e_C\mathcal R\simeq 
\langle h_C+I\rangle$, i.e. there exists a polynomial $g_C$ such that the idempotent $e_C$ is the element $g_C h_C + I$ and  $g_C h_C + I_C = 1 + I_C$. Any ideal of $\mathcal R$ is principally generated by $G+I$ where 
$G=\sum_{i=0}^{t-1}a^iG_{i}$ and $G_{i}$ is a sum of primitive idempotents $e_C$ described before (see \cite[Corollary 3.12]{Serial}).

\begin{remark}
Note that this decomposition includes the  case in \cite{Indian, Habibul} where, in the first reference there are two variables where $R=\mathbb F_q$ and $t_1,t_2$ split completely in linear factors over $\mathbb F_q[X]$ and, in the second one there is one polynomial in $R=\mathbb F_q$ whose roots are all the elements in the field.
\end{remark}

 	\section{Structure of codes over $\mathcal R$}\label{sec:codesovermathcal}

 We will define a \emph{linear code $\mathcal K$ of length $n$ over the ring $\mathcal R$} as an $\mathcal R$-submodule of $\mathcal R^n$. The \emph{Euclidean dual} of $\mathcal K$ will be denoted by $\mathcal K^\perp$ and it is given by the set $\{\mathbf x\in \mathcal R^n\mid \mathbf x\cdot\mathbf k=0\,\hbox{for all } \mathbf k \in\mathcal K \}$, where $\cdot$ is the Euclidean inner product in $\mathcal R^n$.
 	Note that, since $\mathcal R=\sum_{C\in \mathcal C}e_C \mathcal R $,
 	 for each $\mathbf x\in \mathcal R^n$ we can define the projection of  $\mathbf x$ by $e_{C^\prime}$ as $\mathbf x_{C^\prime}=(x_{1,C^\prime}, \ldots , x_{n,C^\prime})\in R^n$ where $x_i=\sum_{C\in\mathcal C} x_{i,C} e_C$ and $x_{i,C}\in R$ for $i=1,\ldots , n$.	
   Indeed,  $\mathbf x_{C^\prime}= \mathbf x\cdot e_{C^\prime}$  for each $C\in \mathcal C$ and for a given linear code $\mathcal K$ of length $n$ over the ring $\mathcal R$ we can define the following codes
 	\begin{equation}\label{eq:dec}
 	    \mathcal K_C = \left\{\mathbf x_C\mid \mathbf x \in \mathcal K\right\},
 	\end{equation}
  where $C$ ranges in the set of classes in $\mathcal C$.
 It is clear that $\mathcal K_C$ is an $R$-linear code  and that 
 $\mathcal K=\bigoplus_{C\in \mathcal C} \mathcal K_C e_C$. Moreover, if $\mathcal K=\bigoplus_{C\in \mathcal C} \mathcal K_C e_C$ then $\mathcal K^\perp=\bigoplus_{C\in \mathcal C} \mathcal K_C^\perp e_C$ (note that we slightly abuse the notation since first orthogonality is in $\mathcal R$ and the second one in $R$).
 
 	\subsection{Constacyclic codes over $\mathcal R$}\label{sec:constacyclic}
 	
 	A \emph{$\lambda$-constacyclic code $\mathcal K$ of
length $n$ over $\mathcal R$} can be regarded as an ideal of $\mathcal R[x]/\langle x^n -\lambda\rangle$ where $\lambda$ is a unit in $ \mathcal R$ (if $\lambda=1$ is called a cyclic code). It is clear that if $\mathcal K$  is a $\lambda$-constacyclic of length $n$ then $\mathcal K^\perp$  is a $\lambda^{-1}$-constacyclic of length $n$. Note that, as was stated above, $\lambda=\sum _{C\in \mathcal C} \lambda_C\cdot e_C$ where $\lambda_C\in R$ and $\lambda$ is a unit if  the $\lambda_C$ is a unit in $R$ for each $C\in \mathcal C$. Henceforth, $\mathcal K$ is $\lambda$-constacyclic of length $n$ in $\mathcal R$  if $\mathcal K_C$ is $\lambda_C$-constacyclic of length $n$ in $R$ for each $C\in \mathcal C$. Then the following result follows and it can be proven in the same fashion as \cite[Theorem 3]{Indian2}.
\begin{proposition}\label{indian}
If $\mathcal K=\bigoplus_{C\in \mathcal C} \mathcal K_C e_C$ is a $\lambda$-constacyclic of length $n$ over $ \mathcal R$, then $\mathcal K^\perp$ is a $\lambda^{-1}$-constacyclic of length $n$  over $\mathcal R$ where  $\mathcal K^\perp=\bigoplus_{C\in \mathcal C} \mathcal K_C^\perp e_C$ and $\lambda_C^{-1}=\lambda_C^\perp$. Furthermore, 
$\mathcal K$ is self-dual if it is $\sum_{C\in \mathcal C} \pm e_C$, i.e. $\lambda^2=1$.
\end{proposition}

The following result characterizes, in a polynomial way, the class of constacyclic codes over $\mathcal R$. First we will introduce the following lemma that characterizes constacyclic codes over chain rings, it can be found in \cite{Salagean} or in a more general way that can be also use for the multivariable case in the language of Canonical Sets of Generators in \cite{repeated}.

\begin{lemma}[\cite{Salagean}]\label{prop:salagean} A non-zero $\lambda_C$-constacyclic code $\mathcal K_C$ over the chain ring $R$ with maximal ideal $\mathfrak m=\langle a\rangle$ and nilpotency index $s$  has a generating set in standard form 
\begin{equation}
S=\{ a^{b_0} g_{b_0}, a^{b_1} g_{b_1} ,\ldots, a^{b_u} g_{b_u} \}  
\end{equation}
such that $\mathcal K_C=\langle S\rangle \triangleleft R[X]/\langle X^n-\lambda_C \rangle$ and
\begin{enumerate}
    \item $0\leq b_0< b_1<\ldots <b_u<t $,
    \item $g_{b_i}$ is a monic polynomial in $R[x]$ for $i=1,\ldots, u$,
    \item $\mathrm{deg}\, g_{b_i}> \mathrm{deg}\, g_{b_{i+1}}$ for $i=1,\ldots, u-1$,
    \item $g_{b_u}|g_{b_{u-1}}|\ldots |g_{b_0}|X^n-\lambda_C$.
\end{enumerate}
Moreover, if $d_i = \mathrm{deg}\, g_{b_i}$ for $i=1,\ldots, u$, then $|\mathcal K_{\lambda_C}|=|R/\mathfrak m|^{\sum_{i=0}^u (s-b_j )(d_{i-1} -d_i )}$ and the code is principal as ideal
$$\mathcal K_C=\Big\langle G_C=\sum_{j=0}^u a^{b_i} g_{b_i}\Big\rangle \triangleleft R[X]/\langle X^n-\lambda_C \rangle.$$
\end{lemma}
\begin{proposition}\label{prop:consta}
Let $\mathcal K=\bigoplus_{C\in \mathcal C} \mathcal K_C e_C$ be a $\lambda$-constacyclic of length $n$ over $\mathcal R$ and 
suppose that the $\lambda_C$-constacyclic codes $\mathcal K_C$ are generated by $G_C(x)$ defined as in the previous Lemma for each $C\in\mathcal C$. Then there exists a polynomial $$\mathcal G(X)=\sum_{C\in \mathcal C} G_C e_C$$
in $\mathcal R[X]$ such that
  $\mathcal K=\Big\langle \mathcal G \Big\rangle \triangleleft \mathcal R[X]/\langle X^n-\lambda\rangle$,    and  $|\mathcal K|=\prod_{C\in\mathcal C}|\mathcal K_{\lambda_C}|$.
 \end{proposition}\begin{proof}
    It is straightforward from the reasoning above.
\end{proof}
	\subsection{Consta-abelian codes over $\mathcal R$}

We can state a similar result as Lemma~\ref{prop:salagean} in the case of abelian codes over finite chain rings in terms of Canonical Set of Generators \cite{Serial,repeated} but, as it was pointed in \cite[Corollary 1]{Principal}, in the case of Abelian codes they are principal if the length of the code is coprime with the characteristic of the chain ring, thus we will restrict ourselves to that case stated in \cite[Sections 4 and 5]{Serial}. In the case of consta-Abelian, that is not the general case (see \cite[Example 1]{Principal} where it is shown that  negacyclic codes over $\mathbb Z_4$ defined by a multiple root polynomial can be seen as a principal ideals). Anyway, since our purpose is defining polyadic codes using splittings of the roots we will restrict to simple root codes. 

Given an Abelian group $A$ expressed as $A=\prod_{i=1}^\delta Z_i$ where, for each $i$, $Z_i$ is a cyclic group  and let $r=\prod_{i=1}^\delta r_i$  where $
r_i$ is the size of the cyclic component $Z_i$ and   $\mathrm{gcd}(r_i,q)=1$ for each $i=1,\ldots, \delta$. 
\begin{definition} 
\emph{An Abelian code over the ring} $\mathcal R$ with underlying group $A=\prod_{i=1}^\delta Z_i$  is an ideal of the ring $\mathcal R[Y_1,\ldots, Y_\delta ]/I_A$, where   $I_A=\langle Y_1^{r_1}-1,\ldots, Y_\delta^{r_\delta }-1  \rangle \subset \mathcal R[Y_1,\ldots, Y_\delta ]$.  Consider now the  ambient space 
\begin{equation}
    \mathcal R_{A,\bm \lambda}=\mathcal R[Y_1,\ldots, Y_\delta ]/I_{A,{\bm \lambda}}= \mathcal R[Y_1,\ldots, Y_\delta ]/\langle Y_1^{r_1}-\lambda_1,\ldots, Y_\delta^{r_\delta }-\lambda_\delta  \rangle, 
\end{equation} where the element $\lambda_i$ is an invertible element in $\mathcal R$ for each $i=1,\ldots, \delta$.
An ideal in $\mathcal R_{A,\bm \lambda}$ is called a ${\bm \lambda}=(\lambda_1,\ldots ,\lambda_\delta)$-\emph{consta-abelian code} with underlying group $A$.
\end{definition}

As in Section~\ref{sec:constacyclic} we can state the following results. The first one is a straightforward generalization of Proposition~\ref{indian}.
\begin{proposition}
If $\mathcal K=\bigoplus_{C\in \mathcal C} \mathcal K_C e_C$ is the decomposition in Equation~(\ref{eq:dec}) of a  $\bm\lambda$ consta-Abelian code with underlying group $A$ over $\mathcal R$, then $\mathcal K^\perp$ is a $\lambda^{-1}=(\lambda_{1}^{-1},\ldots, \lambda_{\delta}^{-1})$ consta-Abelian code with underlying group $A$ over $\mathcal R$, where  $\mathcal K^\perp=\bigoplus_{C\in \mathcal C} \mathcal K_C^\perp e_C$ and $\lambda_C^{-1}=(\lambda_{1,C}^{-1},\ldots, \lambda_{\delta,C}^{-1}) =\lambda_C^\perp$. Furthermore, 
$\mathcal K$ is self-dual if it is  $\sum_{C\in \mathcal C} \pm e_C$, i.e. $\lambda^2=(\lambda_{1}^{2},\ldots, \lambda_{\delta}^{2})=\mathbf 1$.
\end{proposition}
 A particular version of Theorem 3.13~in~\cite{Serial} (and \cite[Corollary 3.14]{Serial} for devising the sizes of the ideals) suited to our setting provides us  the following analogous results to Lemma~\ref{prop:salagean} and Proposition~\ref{prop:consta}.
 
\begin{lemma}A non-zero $\bm \lambda_C$ consta-Abelian  code $\mathcal K_C$ over the chain ring $R$ with maximal ideal $\mathfrak m=\langle a\rangle$ and nilpotency index $s$  has a generating set in standard form 
\begin{equation}
S=\{ a^{b_0} G_{b_0}, a^{b_1} G_{b_1} ,\ldots, a^{b_u} G_{b_u} \} \subset R[Y_1,\ldots, Y_\delta ]
\end{equation}
such that $\mathcal K_C=\langle S\rangle \triangleleft R[Y_1,\ldots, Y_\delta ]$ and
\begin{enumerate}
    \item $0\leq b_0< b_1<\ldots <b_u<t $,
    \item $G_{b_i}$ is a monic polynomial in $R[Y_1,\ldots, Y_\delta ]$ for $i=1,\ldots, u$.
\end{enumerate}
Moreover,  $|\mathcal K_{\lambda_C}|=|R/\mathfrak m|^{\sum_{i=0}^{t-1}(t-i)N_i}$ (where $N_i$ is the number of zeros in $H_1\times \cdots \times ×H_r$ of  $\bar G_{b_i}$) and the code is principal as ideal
$$\mathcal K_C=\Big\langle G_C=\sum_{j=0}^u a^{b_i} G_{b_i}\Big\rangle \triangleleft R[Y_1,\ldots, Y_\delta ]/\langle Y_1^{r_1}-\lambda_1,\ldots, Y_\delta^{r_\delta }-\lambda_\delta  \rangle.$$
\end{lemma}

\begin{proposition}\label{prop:constaabelian}
Let $\mathcal K=\bigoplus_{C\in \mathcal C} \mathcal K_C e_C$ be a ${\bm \lambda}=(\lambda_1,\ldots ,\lambda_\delta)$-{consta-abelian code} with underlying group $A$ of length $n$ over $\mathcal R$ and 
suppose that the $\lambda_C$-constacyclic codes $\mathcal K_C$ are generated by $G_C(x)$ defined as in the previous Lemma for each $C\in\mathcal C$. Then there exists a polynomial  $\mathcal G(X)=\sum_{C\in \mathcal C} G_C e_C$
 such that
 $\mathcal K=\Big\langle \mathcal G \Big\rangle \triangleleft {\mathcal R}_{A,\bm \lambda}$   and  $|\mathcal K|=\prod_{C\in\mathcal C}|\mathcal K_{\lambda_C}|$.
\end{proposition}

\section{Polyadic codes over chain ring}\label{sec:polychain}

\subsection{Splittings and multipliers}\label{sec:spl}
As in the section before, let $A=\prod_{i=1}^\delta Z_i$ be a finite Abelian group, $r=\prod_{i=1}^\delta r_i$   where $
r_i=|Z_i|$ and   $\mathrm{gcd}(r_i,q)=1$ for each $i=1,\ldots, \delta$. We will associate to $A$ the ideal  $I_A=\langle Y_1^{r_1}-1,\ldots, Y_\delta^{r_\delta }-1  \rangle \subset R[Y_1,\ldots, Y_\delta ]$ and clearly $R[A]\simeq R[Y_1,\ldots, Y_\delta ]/I_A$. We will denote as $\mathcal C_A$ the set of  the cyclotomic classes associated to $I_A$ as in Section~\ref{sec:pre}.  We will define a \textit{splitting} of $A$ following the notation in \cite{Ling1}. For that, we will consider the commutative group $A_{\star}=(A,\star)$ given by the component-wise multiplication $\star$ in $A$ derived from the multiplication in the components $Z_i\simeq \mathbb Z/\mathbb Z_{r_i}$ and $A_{\star}^*$ its group of units. Any $u=(u_1,\ldots , u_\delta)\in A_{\star}^*$ defines an action $u_\star$ over $A$ given by $a=(a_1,\ldots , a_\delta)\mapsto u_\star(a)=(u_1a_1,\ldots , u_\delta a_\delta)$ for all $a $ in $A$ . We extend this concept to a union $C$ of cyclotomic classes in $\mathcal C_A$ defining $u_\star(C)$ as the union of the exponents of the images of the elements $u_\star(a)$ where $a\in A$ is associated to  an element in $C$. We call this $u_\star$ a \textit{multiplier}.

\begin{definition}
For a positive integer $m \geq 2 $ and a nonempty set $S_\infty \subset A$, an $m-$ splitting of $A$ is a $m$-tuple $\mathcal S=(S_\infty, S_0, S_1, ..., S_{m-1})$ which satisfies the follow conditions:
\begin{enumerate}
    \item Each set $S_\infty, S_0, S_1, ..., S_{m-1}$ is a union of  cyclotomic classes in $\mathcal C_A$,
    \item The sets $S_\infty, S_0, S_1, ..., S_{m-1}$ are disjoint and form a partition of $\mathcal C_A$,
    \item There exists $u \in A_{\star}^*$ such that $u_*(S_\infty)=S_\infty$ and $u_\star(S_i)=S_{i+1}$ where $u_\star$ is a multiplier.
\end{enumerate}
\end{definition}
In the literature, usually, the splittings are defined over cyclotomic cosets of modular integers, note that for this case there is a one-to-one correspondence with our classes of roots.
It is clear that in a splitting the class given by  $\{\mathbf 0 \}$ is contained in the set $S_\infty$. 

For a given set $S$ which is a union of cyclotomic classes and for  a chain ring $R$ with quotient field $\mathbb F_q$ we will denote  the ideal on $R[Y_1,\ldots, Y_\delta]/I_A$ by $I_S$ the ideal given by $I_S=\bigcap_{C\in\mathcal C_A,C\in S} I_{C}$. Note that in the case that $R$ is a finite field, $I_S$ denotes 
 the polynomial ideal in $R[Y_1,\ldots, Y_\delta ]/I_A$ whose elements vanish when evaluated in all the elements in $S$.
%For any $X \subset G$, the set$ I_X=\{ c\in F[G] : \hat{c_x}=0 $ for  all $x \in X \}$ is an ideal in $F[G]$. 
\subsection{Polyadic abelian codes over chain rings}

\begin{definition}[Polyadic Abelian codes over a chain ring] Let $ R$ be a chain ring.
Let $A$ be a finite Abelian group, $0\leq i \leq m-1$, $\mathcal S=(S_\infty, S_0, S_1, ..., S_{m-1})$ a $m$-splitting of the cyclotomic classes  $\mathcal C_A$ associated to  $I_A$ and  $S'_{\infty}=S_{\infty}\setminus\{\mathbf 0\}$. The ideals (codes)
\begin{equation}
    K_i=I_{(S'_{\infty}\cup S_i)^c}, \widehat{K_i}=I_{S'_{\infty}\cup S_i}, L_i=I_{S_{\infty}\cup S_i}, \widehat{L_i}=I_{(S_{\infty}\cup S_i)^c}
\end{equation}
defined in $R[Y_1,\ldots, Y_\delta]/I_A$ are called\textit{ polyadic codes}. $K_i$ and $\widehat {K_i}$ are called even-like codes  and  $L_i$, $\widehat {L_i}$ are called odd-like codes.
\end{definition}

The following result follows directly from the definition (see \cite[Theorem 2.1]{Ling1}) for its counterpart of cyclic codes over finite fields).
\begin{proposition}\label{pro:dec_spl}
For $i\neq j$, $i,j\in\{0,1,\ldots, m-1\}$
\begin{itemize}
\item The following identities hold
\begin{enumerate}
    \item $K_i\cap K_j=I_{{S^\prime_\infty}^c}$ and $K_0+K_1+\ldots + K_{m-1}=I_{ \{\mathbf 0\}}.$
     \item $\widehat{ K_i} +\widehat { K_j}=I_{{S^\prime_\infty}}$ and $\widehat {K_0}\cap \widehat{K_1}\cap \ldots \cap\widehat  {K_{m-1}}=I_{\{\mathbf 0\}^c}.$
     \item $L_i +L_j=I_{{S_\infty}}$ and $L_0\cap L_1\cap \ldots \cap L_{m-1}=\{ \mathbf 0\}.$
      \item $\widehat{L_i}\cup\widehat{L_j}=I_{{S_\infty}^c}$ and $\widehat{L_0}+\widehat{L_1}+\ldots + \widehat{ L_{m-1}}=R[Y_1,\ldots, Y_\delta]/I_A.$
\end{enumerate}
\item $K_i+\widehat{K_i}= R[Y_1,\ldots, Y_\delta]/I_A=L_i+\widehat{L_i}$.
\item For $0\leq i \leq m-1$, all the codes $K_i$ are equivalent codes.  The same is true for the other families of codes $\widehat{K_i}$, $L_i$, and $\widehat{L_i}$.
\end{itemize}
\end{proposition}
Note that the last fact  is a straightforward result of  multiplier $u_*$ having the  property $u_*(S_i)=S_{i+1}$. Since each $K_i $ is uniquely determined by the $m$-splitting set $S_i $ it can be regarded as a permutation thus  each code $K_i $ is equivalent to the other $K_{i+1} $.

\begin{proposition}
For an m-adic code $K$ over a finite chain ring $R$, let $\widetilde{K}=K\cdot I_{{S_\infty}}$ as ideals in $ R[Y_1,\ldots, Y_\delta ]/I_A$. Then, for all $i\in\{0,1,\ldots, m-1\}$ we have 
$$ \widetilde{K_i}=\widetilde{\widehat{L_i}}\quad \hbox{ and } \quad \widetilde{\widehat{K_i}}=\widetilde{ L_i}=L_i.$$
\end{proposition}
 The code $\widetilde{K}=I_{{S_\infty}}$ in the proposition above is called the \textit{even-like} subcode of $K$ when $S_\infty=\{ 0\}$ and the codewords in $K\setminus \widetilde{K}$ are called \textit{odd-like} in that case.
 
For $0\leq i \leq m-1$, let $\bar{e_i}$ and $\bar{e'_i}$ be the even-like idempotent generators of even-like codes $K_i$ and $\widehat {K_i}$ respectively, $\bar{d_i}$ and $\bar{d'_i}$ be the odd-like idempotent generators of even-like codes $L_i$ and $\widehat {L_i}$ respectively in $R[Y_1,\ldots, Y_\delta]/I_A$, respectively given in Section~\ref{sec:pre}.

 If we take the element $\sigma=-1\in A_\star$ it is clear that it induces a permutation of the cyclotomic cosets but it could be the case that it does not induce a permutation on the sets $S_0,\ldots, S_{m-1}$ of an $m$-splitting. The following result follows directly form the finite field case in \cite[Proposition 2.2]{Ling1} and the characterization of the dual of an Abelian code over a chain ring in \cite[Sections 4 \& 5]{Serial}.
 
 \begin{proposition}
 Suppose that $\sigma_\star(S_\infty)=S_\infty$ and that $\sigma_\star$ is a permutation of $S_0,\ldots , S_{m-1}$ such that $\sigma_\star(X_i)=X_{\Tilde{\sigma}(i)}$, for $i\in\{1,\ldots , m-1\}$. Then
 $$K^\perp_i=\widehat{K}_{\Tilde{\sigma}(i)} \quad \hbox{ and }\quad L^\perp_i=\widehat{L}_{\Tilde{\sigma}(i)}. $$
 \end{proposition}
 
 \begin{remark}
Note that the existence of polyadic abelian codes over the finite chain ring $R$ relies on the existence of polyadic abelian codes over $\mathbb F_q= R/\mathfrak m $ since we used the same cyclotomic clases. We refer to \cite[Section III]{Ling1} for that study.
\end{remark}
 
 \subsection{Polyadic Consta-Abelian codes over chain rings}\label{sec:polyconsta}
 
 We will follow mainly the ideas in \cite{Lim2005} for describing consta-Abelian serial codes over a chain ring $R$ defined by the Abelian group $A=\prod_{i=1}^\delta Z_i$ and   ${\bm \lambda}=(\lambda_1,\ldots ,\lambda_\delta)\in (R^*)^\delta$.  The ambient space is given by $R_{A,\bm \lambda}= R[Y_1,\ldots, Y_\delta ]/I_{A,{\bm \lambda}}=  R[Y_1,\ldots, Y_\delta ]/\langle Y_1^{r_1}-\lambda_1,\ldots, Y_\delta^{r_\delta }-\lambda_\delta  \rangle $. If we consider the set of roots of the polynomials $\overline{ Y_1^{r_1}-\lambda_1},\ldots, \overline{Y_\delta^{r_\delta }-\lambda_\delta}$ in a extension field of $R/\mathfrak m$ they are given by $(\beta_1\xi_1^{i_1},\ldots , \beta_\delta\xi_\delta^{i_\delta} )$ where $\beta_j$ is a primitive $r_j$-th root of $\bar\lambda_j$,  $\xi_j$ is a primitive $r_j$-th root of unity  and $0\leq i_j \leq n-1$ for $j=1,\ldots , \delta$.
 
 As in the section above, we will consider the commutative group $A_{\star}=(A,\star)$. For each set $S\subset A$ one can define $\bar S=\{1+r(s-1)\mid s\in S\}$. We say that $S\subset A$ defines an orbit with respect to $r$ if $\bar S$ is a cyclotomic coset of $A$, in other words, defines the exponents of a cyclotomic class of the associated Abelian code $R[A]$ that we will denote as $C_{\bar S}$.  
 \begin{definition} \label{def:classes} Let $\Theta$ be a union of orbits in $A$ w.r.t.  $r$. 
For a positive integer $m \geq 2 $ and a   set $S_\infty \subset \Theta$, an $m-$ splitting of $\Theta$ w.r.t. $r$  is a $m$-tuple $\mathcal S=(S_\infty, S_0, S_1, ..., S_{m-1})$ which satisfies the follow conditions:
\begin{enumerate}
    \item Each set $S_\infty, S_0, S_1, ..., S_{m-1}$ is a union of orbits in $A$ w.r.t. $r$,
    \item The sets $S_\infty, S_0, S_1, ..., S_{m-1}$ are disjoint and form a partition of $\Theta$,
    \item There exists $u \in A_{\star}^*$ such that $u_*(C_{\bar S_\infty})=C_{\bar S_\infty}$ and $u_*(C_{\bar S_i})=C_{\bar S_{i+1}}$ where $u_*$ is a multiplier.
\end{enumerate}
\end{definition}
We say that the splitting is non-trivial if $S_\infty\subsetneq \Theta$, that is $S_i\neq \emptyset$ for $i=0,\ldots ,n-1$.

Given a splitting of $\Theta$ w.r.t. $r$ as above, the ideals (codes) given by its defining sets
\begin{equation}
    K_i=I_{(  \bar S_i)^c}, \widehat{K_i}=I_{  \bar S_i}, L_i=I_{\bar S_{\infty}\cup \bar S_i}, \widehat{L_i}=I_{(\bar S_{\infty}\cup \bar S_i)^c}
\end{equation}
defined in $R[Y_1,\ldots, Y_\delta]/I_A$ are called\textit{ polyadic codes}. $K_i$ and $\widehat {K_i}$ are called even-like codes  and  $L_i$, $\widehat {L_i}$ are called odd-like codes.

Note that in the case of $A$ being a cyclic group and $\Theta=A$ we are in the case of splittings for constacyclic codes, moreover in that case if $S_\infty=\emptyset$ they are called Type I, otherwise they are called Type II (See for example \cite{Chen}).  Now we have a similar result to Proposition~\ref{pro:dec_spl} that can be stated for chain rings also (see \cite[Theorem 7.2]{Lim2005} for the finite field case).

\begin{proposition} 
For $i\neq j$, $i,j\in\{0,1,\ldots, m-1\}$
\begin{itemize}
\item The following identities hold
\begin{enumerate}
     \item $L_i +L_j=I_{{S_\infty}}$ and $L_0\cap L_1\cap \ldots \cap L_{m-1}=\{ \mathbf 0\}.$
      \item $\widehat{L_i}\cup\widehat{L_j}=I_{{S_\infty}^c}$ and $\widehat{L_0}+\widehat{L_1}+\ldots + \widehat{ L_{m-1}}=R[Y_1,\ldots, Y_\delta]/I_{A,{\bm \lambda}}.$
\end{enumerate}
\item $  R[Y_1,\ldots, Y_\delta]/I_{A,{\bm \lambda}}=L_i+\widehat{L_i}$.

 \item For $0\leq i \leq m-1$, all the codes $L_i$ are equivalent codes.  The same is true for the  family of codes  $\widehat{L_i}$.
\end{itemize}
\end{proposition}
 
\section{Polyadic abelian codes over serial rings }\label{sec:core1}
In this section we describe polyadic codes of length $n$ over the ring $\mathcal{R}$ expressed as in Equation~\ref{eq:curlyR} by using its primitive idempotents $e_C$ in the Section~\ref{sec:intro} where $C\in \mathcal C$ is identified. Let us divide the classes of $\mathcal C$ from $\{1,\ldots, |\mathcal C|\}$ into $m$ disjoint sets and denote these disjoint sets with $A_i$ for $i=1,\ldots,m$. $\mathcal{C}$ can be written as follows:
\begin{equation} \label{eq:mpartition}
    \mathcal C=\{ C_i \mid i=1,\ldots, |\mathcal C|\}= A_1\cup \dots\cup A_m
\end{equation} Consider the condition that  each  $A_i$ is non-empty set  if $|\mathcal{C}|\geq m$. Moreover, in this case $|A_i|=t_i$, $1\leq t_i \leq |\mathcal{C}|-m+1$. Otherwise, $|\mathcal{C}|$ sets in the partition in the Equation~\ref{eq:mpartition} are non-empty set has only one element and the remaining $m-|\mathcal{C}|$ are empty. So in this case, $|A_i|=t_i=1$ if $A_i$ is non-empty and $|A_i|=t_i=0$ if $A_i$ is empty. It can be easily seen that $|\mathcal{C}|=\sum_{i=1}^{m}t_i$.

We define $\theta_{A_i}=\sum_{C_j\in A_i} e_{C_j}$ for each $i=1,\ldots, m$. Assume that $\theta_{A_i}=0$ when $A_i=\emptyset$. It is easily seen that
$\sum_{i=1}^{m}\theta_{A_i}=\sum e_{C}=1_{\mathcal R}$, $\theta_{A_i}^2=\theta_{A_i}$ and $\theta_{A_i}.\theta_{A_j}=0$ for all $i \neq j$. Note also that this idempotents  in $\mathcal R$ seen as generators correspond to a disjoint decomposition of $\mathcal R$ as a sum of serial codes. 

Let $E_i,E_i',D_i,D_i'$ be the idempotent generators of polyadic codes over $R[A]$ defined as in the Section~\ref{sec:spl} for  $i=1,\ldots,m$. 
$E_i$ and $E_i'$'s are even-like ones, while the others are odd-like ones.

From now on, we can define the idempotents to obtain polyadic codes over the serial ring $\mathcal R[A]=R[X_1,\ldots, X_s,Y_1,\ldots, Y_\delta]/\langle I  , I_A\rangle$. These idempotents can be written as follows: 

Note that the following index number $k_{ij}$ we will use in order to enumerate all idempotents is the smallest positive integer which is equivalent to  the number $i-j+1$ (i.e $k_{ij}=i-j+1 $ $mod \  (m)$) and the structure of the positive integer $k_{ij}$ forces the cyclicity of the mapping $u_\star$ over the new idempotents.

\begin{itemize}

    \item Odd-like idempotent generators over the ring $\mathcal{R}[A] $ for each $j=2,\ldots, m$
    \begin{itemize}
        \item $\mathcal{D}_1=\sum_{i=1}^{m} \theta_{A_i}D_i $ where $D_i$ is the idempotent generator for $L_i$.
        \item $\mathcal{D}_j=u_*(\mathcal{D}_{j-1})=\sum_{i=1}^{m} \theta_{A_i}D_{k_{ij}} $
        \item $\mathcal{D}_1'=\sum_{i=1}^{m} \theta_{A_i}D_i' $ where $D_i^\prime$ is the idempotent generator for $\widehat{L_i}$.
        \item $\mathcal{D}_j'=u_*(\mathcal{D}_{j-1}')=\sum_{i=1}^{m} \theta_{A_i}D'_{k_{ij}} $ 
    \end{itemize}
    \item Even-like idempotent generators over the ring $\mathcal{R}[A]$ for each $j=2,\ldots, m$
    \begin{itemize}
        \item  $\mathcal{E}_1=\sum_{i=1}^{m} \theta_{A_i}E_i $ where $E_i$ is the idempotent generator for $K_i$.
        \item $\mathcal{E}_j=u_*(\mathcal{E}_{j-1})=\sum_{i=1}^{m} \theta_{A_i}E_{k_{ij}} $ 
        \item $\mathcal{E}_1'=\sum_{i=1}^{m} \theta_{A_i}E_i' $ where $E_i$ is the idempotent generator for $\widehat{K_i}$.
        \item $\mathcal{E}_j'=u_*(\mathcal{E}_{j-1}')=\sum_{i=1}^{m} \theta_{A_i}E'_{k_{ij}} $ 
    \end{itemize}
\end{itemize}
 %Let $\mathcal{D}_1=\sum_{i=1}^{m} \theta_{A_i}D_i $, $\mathcal{D}_j=u_*(\mathcal{D}_{j-1})=\sum_{i=1}^{m} \theta_{A_i}D_{k_{ij}} $, $\mathcal{D}_1'=\sum_{i=1}^{m} \theta_{A_i}D_i' $ and $\mathcal{D}_j'=u_*(\mathcal{D}_{j-1}')=\sum_{i=1}^{m} \theta_{A_i}D'_{k_{ij}} $ where $k_{ij}=m+i-j+1 $ $mod \  (m)$ for each $j=2,\ldots, m$ be odd-like idempotents in the ring $\mathcal{R}[x]/\langle x^n-1 \rangle $. 
 %Similarly, let   $\mathcal{E}_1=\sum_{i=1}^{m} \theta_{A_i}E_i $, $\mathcal{E}_j=u_*(\mathcal{E}_{j-1})=\sum_{i=1}^{m} \theta_{A_i}E_{k_{ij}} $, $\mathcal{E}_1'=\sum_{i=1}^{m} \theta_{A_i}E_i' $ and $\mathcal{E}_j'=u_*(\mathcal{E}_{j-1}')=\sum_{i=1}^{m} \theta_{A_i}E'_{k_{ij}} $ where $k_{ij}=m+i-j+1 $ $mod \  (m)$ for each $j=2,\ldots, m$ be even-like idempotents in the ring $\mathcal{R}[x]/\langle x^n-1 \rangle $. 

 Therefore, we can obtain odd-like (or even-like) polyadic codes over $\mathcal{R}$ by using the odd-like idempotents $\mathcal{D}_j$ and $\mathcal{D}_j'$ (or even-like idempotents $\mathcal{E}_j$ and $\mathcal{E}_j'$) for $j=1, \ldots ,m$. Let the polyadic codes associated with the idempotents $\mathcal{D}_j$, $\mathcal{D}_j'$, $\mathcal{E}_j$ and $\mathcal{E}_j'$ over $\mathcal{R}$ be called as $\mathcal{P}_j$, $\widehat{\mathcal{P}_j}$, $\mathcal{Q}_j$ and $\widehat{\mathcal{Q}_j}$, respectively. So, the desired polyadic codes are generated by the idempotents such that $\mathcal{P}_j=\langle \mathcal{D}_j \rangle $, $\widehat{\mathcal{P}_j}=\langle \mathcal{D}_j' \rangle $, $\mathcal{Q}_j=\langle \mathcal{E}_j \rangle $ and $\widehat{\mathcal{Q
 }_j}=\langle \mathcal{E}_j' \rangle $. 
 
 \begin{remark}
   By choosing $A_i$ once, we now get a new idempotent to define a polyadic code (odd-like or even-like) and one can easily see that the polyadic codes obtained by taking image of these idempotents under the multiplier $u_*$ are all equivalent codes. But if we change our choice for $A_i$ we will get a new odd-like (or even-like) polyadic code that is inequivalent code to the other codes corresponding to other selections. So we can count the number of these inequivalent codes as in the follow theorem.
 \end{remark}

\begin{theorem} [Number of polyadic codes] \label{theo:number} The following statements are hold:

\begin{enumerate}
    \item If  $|\mathcal{C}| \geq m $, then the number of inequivalent odd-like (or even-like) polyadic codes over the ring $\mathcal{R}$ is equal to { $$ \frac{2}{m} \sum_{t_{m-1}=1}^{|\mathcal{C}|-T_{m-2}-1} \ldots \sum_{t_{2}=1}^{|\mathcal{C}|-T_1-(m-2)}\sum_{t_{1}=1}^{|\mathcal{C}|-(m-1)} \binom{|\mathcal{C}|}{t_1} \binom{|\mathcal{C}|-T_1}{t_2} \ldots \binom{|\mathcal{C}|-T_{m-2}}{t_{m-1}}  , $$}
    where $T_i=\sum_{j=1}^{i} t_j$.
    \item If $|\mathcal{C}| < m $, then the number of inequivalent odd-like (or even-like) polyadic codes over the ring $\mathcal{R}$ is equal to
    $$ \frac{2}{m} (|\mathcal{C}|)!\binom{m}{|\mathcal{C}|}  $$
    
\end{enumerate}
\end{theorem}

\begin{proof}
    First, we will prove the desired number in the case $|\mathcal{C}| \geq m $. To count the odd-like polyadic codes, we should calculate the number of idempotent generators by using the equalities $\mathcal{D}_j=u_*(\mathcal{D}_{j-1})=\sum_{i=1}^{m} \theta_{A_i}D_{k_{ij}} $ and $\mathcal{D}_1=\sum_{i=1}^{m} \theta_{A_i}D_i $ for $j=2,\ldots, m$. The total number of odd-like idempotent generators obtained by these equations  is determined by the number of choices of $\theta_{A_i}$ for $i=1,\ldots, m$. Recall that $|\theta_{A_i}|=t_i$ and the total number of idempotents of type $e_C$ is $|\mathcal{C}|$.  $\theta_{A_1}$ can be chosen in $\binom{|\mathcal{C}|}{t_1}$ different ways out of $|\mathcal{C}|$ idempotents. $\theta_{A_2}$ can be chosen in $\binom{|\mathcal{C}|-t_1}{t_2}$ different ways out of $|\mathcal{C}|-t_1$ remained idempotents. In the same fashion, $\theta_{A_{m-1}}$ can be chosen in $\binom{|\mathcal{C}|-(t_1+t_2+...+t_{m-2})}{t_{m-1}}$ different ways and finally $\theta_{A_{m}}$ has to be taken in only one way. Thus, the total number of odd-like idempotent generators is equal to
{$$  \sum_{t_{m-1}=1}^{|\mathcal{C}|-T_{m-2}-1} \ldots \sum_{t_{2}=1}^{|\mathcal{C}|-T_1-(m-2)}\sum_{t_{1}=1}^{|\mathcal{C}|-(m-1)} \binom{|\mathcal{C}|}{t_1} \binom{|\mathcal{C}|-T_1}{t_2} \ldots \binom{|\mathcal{C}|-T_{m-2}}{t_{m-1}}.$$} Since $\mathcal{D}_1=\sum_{i=1}^{m} \theta_{A_i}D_i $ , $\mathcal{D}_j=u_*(\mathcal{D}_{j-1})=\sum_{i=1}^{m} \theta_{A_i}D_{k_{ij}} $ for $j=2,\ldots, m$  and the fact that  $u_*$ only permutes the idempotents $D_i$, the number of idempotent generators obtained $\mathcal{D}_j$ is equal for $j=1,\ldots, m$. So, only one $\mathcal{D}_j$ generates $$ \frac{1}{m} \sum_{t_{m-1}=1}^{|\mathcal{C}|-T_{m-2}-1} \ldots \sum_{t_{2}=1}^{|\mathcal{C}|-T_1-(m-2)}\sum_{t_{1}=1}^{|\mathcal{C}|-(m-1)} \binom{|\mathcal{C}|}{t_1} \binom{|\mathcal{C}|-T_1}{t_2} \ldots \binom{|\mathcal{C}|-T_{m-2}}{t_{m-1}}    $$ odd-like idempotent generators and similarly the number of idempotent generators generated by $\mathcal{D}_j'$ is same. Therefore, we have just counted the number of odd-like idempotent generators (i.e. the number of odd-like inequivalent polyadic codes) as the number stated in the expression of the theorem. The same quantity is valid for the number of even-like inequivalent polyadic codes.
Finally, consider the case $|\mathcal{C}| < m $. Since all non-empty $A_i$ has only one element, the total number of choices $A_i$ is equal to $\binom{m}{|\mathcal{C}|}$ and the number of their displacement is $(|\mathcal{C}|)!$, the total number of odd-like idempotent generators is $(|\mathcal{C}|)!\binom{m}{|\mathcal{C}|}$. The number of inequivalent ones is $\frac{1}{m}(|\mathcal{C}|)!\binom{m} {|\mathcal{C}|}$ by using the same argument as the previous case. Finally by considering the equality of arrivals from both $\mathcal{D}_j$ and $\mathcal{D}_j'$, the total number of inequivalent odd-like polyadic codes is $\frac{2}{m}(|\mathcal{C}|)!\binom{m} {|\mathcal{C}|}$. Similar operations in both cases can be done to compute the number of inequivalent even-like polyadic codes.
\end{proof}

 We will denote by $\mathrm{Rep}(n)$ \emph{the repetition code of length $n$}, that is, the code generated by the polynomial $\sum_{\bm i}Y^{\bm i}\in \mathcal R[A]$ (i.e. the polynomial with all ones as coefficients),  and, as usual,  the \emph{even weight code} is just $\mathrm{Rep}(n)^\perp$. The following two theorems extend those in \cite{Indian2} to affine algebra rings and Abelian codes. 

\begin{theorem} \label{forproof}
    Let $B$ be a subset of $\{1,2,\ldots,m\}$ with at least two elements.  The following propositions are satisfied for the polyadic codes $\mathcal{P}_i$, $\mathcal{Q}_i$ over $\mathcal{R}$ defined as above.  \begin{enumerate}
        \item $\bigcap_{i=1}^m\mathcal{P}_i=\mathrm{Rep}(n)$, the repetition code over $\mathcal{R}$.
        \item $\sum_{i=1}^m\mathcal{P}_i=\sum_{j\in B}\mathcal{P}_j$.
        \item $\bigcap_{i=1}^m\mathcal{Q}_i=\bigcap_{j\in B}\mathcal{Q}_j$
        \item $\sum_{i=1}^m\mathcal{Q}_i=\mathrm{Rep}(n)^\perp$, the even weight code over $\mathcal{R}$.
        \item $\mathcal{Q}_i\cap \mathrm{Rep}(n) =\{0\}$ and $\mathcal{P}_i\cap \mathrm{Rep}(n) =\mathrm{Rep}(n)$ for $ 1\leq i \leq m$.
        \item $\mathcal{P}_i+\mathcal{Q}_i=\mathcal{R}[A] $ and  $\mathcal{P}_i\cap \mathcal{Q}_i=\{0\}$ for $ 1\leq i \leq m$.
    \end{enumerate}
    If we consider $\widehat{\mathcal{P}_i}$'s and  $\widehat{\mathcal{Q}_i}$'s instead of  $\mathcal{P}_i$'s  and $\mathcal{Q}_i$'s  respectively, the previous statements also hold.
\end{theorem}
\begin{proof}
    For the proof of the first statement of the theorem, recall that for any Abelian codes $C$ and $D$, the defining set of the Abelian code $C\cap D$ is the union of the defining set of $C$ and $D$. So the union of all defining sets of $\mathcal{P}_i$ generates the repetition code. Recall also that for any two Abelian codes $C$ and $D$ whose idempotent generators of these codes are $E_1$ and $E_2$ respectively, the idempotent generators of $C\cap D$ and  $C+D$ are $E_1E_2$ and $E_1+E_2-E_1E_2$ respectively. By generalizing of these properties we can obtain the second statement of the theorem. By using that the idempotent generators of $\mathcal{P}_i$ are $\mathcal{D}_i$ as above and the fact that the basic properties of idempotents we get 
    the following expression 
    
 $$\mathcal{D}_1\mathcal{D}_2=\mathcal{D}_1u_*(\mathcal{D}_{1})=\sum_{i=1}^{m} \theta_{A_i}D_i\sum_{i=1}^{m} \theta_{A_i}D_{k_{i2}}=\sum_{i=1}^{m}\theta_{A_i}D_iD_{m-i+1}$$ $$=\sum_{i=1}^{m}\left(\theta_{A_i}\prod_{j=1}^mD_j\right)=\left( \sum_{i=1}^{m}\theta_{A_i}\right)\left(\prod_{j=1}^mD_j\right)=\prod_{j=1}^mD_j$$
By applying similar operations we obtain the following equality $$ \prod_{j \in B} \mathcal{D}_j=\prod_{j=1}^mD_j$$ we can get the 3) in the theorem. The expression in item  4) can easily be obtained by considering 1).

Now, Let the idempotent generator of the $\mathrm{Rep}(n)$ be $i_{\mathrm{Rep}}$. By using that the idempotent generators of $\mathcal{Q}_i$ are $\mathcal{E}_i$ as above and the fact that $\mathcal{E}_i \cdot i_{\mathrm{Rep}}=0$, we get that $\mathcal{Q}_i\cap \mathrm{Rep}(n)={0}$. Now using the equality $\mathcal{D}_i \cdot i_{\mathrm{Rep}}=(1-\mathcal{E}_i)\cdot i_{\mathrm{Rep}} = i_{\mathrm{Rep}}-\mathcal{E}_i \cdot i_{\mathrm{Rep}}$ and the previous result, the equality $\mathcal{D}_i\cdot i_{\mathrm{Rep}} =i_{\mathrm{Rep}}$ arises. Therefore, $\mathcal{P}_i\cap \mathrm{Rep}(n)=\mathrm{Rep}(n)$. Both of these two results give item 5) in the theorem.  
For  proving the last properties in the theorem we will take $i=1$ for convenience. Now  we should find out what are the idempotent generators of the codes $\mathcal{P}_1+\mathcal{Q}_1$ and $\mathcal{P}_1 \cap \mathcal{Q}_1$. Note that the  idempotent generators of the code $\mathcal{P}_1+\mathcal{Q}_1$ and $\mathcal{P}_1 \cap \mathcal{Q}_1$ can be written as $\mathcal{D}_1+\mathcal{E}_1-\mathcal{D}_1\mathcal{E}_1$ and $\mathcal{D}_1\mathcal{E}_1$, respectively.
Since $\mathcal{D}_1\mathcal{E}_1=\sum_{i=1}^{m} \theta_{A_i}D_iE_i=0$ and $\mathcal{D}_1+\mathcal{E}_1=\sum_{i=1}^{m} \theta_{A_i}(D_i+E_i)=\sum_{i=1}^{m} \theta_{A_i}=1$, we get that $ \mathcal{P}_1 \cap \mathcal{Q}_1=\langle \mathcal{D}_i\mathcal{E}_i \rangle=\{0\}$ and $\mathcal{P}_1 + \mathcal{Q}_1= \langle \mathcal{D}_i+\mathcal{E}_i-\mathcal{D}_i\mathcal{E}_i \rangle=\langle 1 \rangle$.
    \end{proof}

\begin{theorem}
    Let $B$ be a subset of $\{1,2,\ldots,m\}$ with at least two elements.  The following statements are satisfied by the polyadic codes $\mathcal{P}_i$ and $\mathcal{Q}_i$ over $\mathcal{R}$ defined as above. 
    \begin{enumerate}
        \item $\mathcal{Q}_i+\mathrm{Rep}(n)=\widehat{\mathcal{P}_i}$ and $\widehat{\mathcal{Q}_i}+\mathrm{Rep}(n)=\mathcal{P}_i$ 
        \item $\mathcal{Q}_i \cap \widehat{\mathcal{Q}_i}=\mathrm{Rep}(n)^\perp$ and $\mathcal{Q}_i \cap \widehat{\mathcal{Q}_i}=\{ 0 \}$
        \item  $\mathcal{P}_i+\widehat{\mathcal{P}_i}=\mathcal{R}[A] $ and $\mathcal{P}_i \cap \widehat{\mathcal{P}_i}=\mathrm{Rep}(n)$

    \end{enumerate}
\end{theorem}

\begin{proof}
  Let  $\mathcal{E}_i$ and $i_{\mathrm{Rep}}$ be the generator idempotents of $\mathcal{Q}_i$ and $\mathrm{Rep}(n)$ respectively. Then  the idempotent generator of the code $\mathcal{Q}_i+\mathrm{Rep}(n)$ can be written as  
   \begin{equation*}
   \begin{split}
   \mathcal{E}_i+ i_{\mathrm{Rep}} -\mathcal{E}_i \cdot i_{\mathrm{Rep}} &=\mathcal{E}_i+ i_{\mathrm{Rep}}=\sum_{j=1}^{m} \theta_{A_j}E_j+ i_{\mathrm{Rep}}\\
   &=\sum_{j=1}^{m} \theta_{A_j}E_j+ i_{\mathrm{Rep}} \cdot \sum_{j=1}^{m} \theta_{A_j}\\
   &=\sum_{j=1}^{m} \theta_{A_j}(E_j+ i_{\mathrm{Rep}})=\sum_{j=1}^{m} \theta_{A_j}D_j'=\mathcal{D}_j'.
   \end{split}
   \end{equation*}
Hence $\mathcal{Q}_i+\mathrm{Rep}(n)=\widehat{\mathcal{P}_i}$ is satisfied. The remaining parts of the result can be proven in a similar way. 
\end{proof}
 Consider $u=(-1,\ldots, -1)=-\mathbf 1$, then $u_\star$   given by $a=(a_1,\ldots , a_\delta)\mapsto u_\star(a)=(-a_1,\ldots , -a_\delta)$ for all $a $ in $A$ and we will denote it by $-\mathbf 1_{\star}$. Then, the following theorem for the existence of Linear Complementary Dual (LCD) codes holds.

\begin{theorem} [LCD codes] \label{the:LCD1}
    Let $\mathcal{Q}_i $ and $\widehat{\mathcal{Q}_i} $ be a pair of even-like polyadic codes with the associated odd-like polyadic codes $\mathcal{P}_i $ and $\widehat{\mathcal{P}_i} $ over the ring $\mathcal{R}$ for    $ 1\leq i \leq m$. The following statements hold. 
    \begin{enumerate}
        \item $\mathcal{Q}_i^{\perp}=-\mathbf 1_{\star}(\mathcal{P}_i) $ and $\widehat{\mathcal{Q}_i}^{\perp}=-\mathbf 1_{\star}(\widehat{\mathcal{P}_i}) $
        \item If $-\mathbf 1_{\star}(E_i)=E_i$  then $\mathcal{Q}_i^{\perp}=\mathcal{P}_i$, $\widehat{\mathcal{Q}_i}^{\perp}=\widehat{\mathcal{P}_i}$ and $\mathcal{Q}_i, \widehat{\mathcal{Q}_i}, \mathcal{P}_i,\widehat{\mathcal{P}_i}$ are LCD codes over $\mathcal{R}$, for $ 1\leq i \leq m$.
    \end{enumerate}

\end{theorem}

    \begin{proof} Without loss of generality we will assume that $i=1$.
       The idempotent generator of  $\mathcal{Q}_1^{\perp}$ is  given by 
       \begin{equation*}
           \begin{split}
               1-\left(-\mathbf 1_{\star}\right)(\mathcal{E}_1) &=\sum_{j=1}^m\theta_{A_j}-\left(-\mathbf 1_{\star}\right)\left(\sum_{j=1}^m\theta_{A_j}E_j\right) =\sum_{j=1}^m\left[\theta_{A_j}(1-\left(-\mathbf 1_{\star}\right)(E_j))\right] \\
               & =\left(-\mathbf 1_{\star}\right)\left( \sum_{j=1}^m \theta_{A_j}D_j\right)=\left(-\mathbf 1_{\star}\right)(\mathcal{D}_1).
           \end{split}
       \end{equation*}
       Hence, $\mathcal{Q}_1^{\perp}=\langle 1-\left(-\mathbf 1_{\star}\right)(\mathcal{E}_1) \rangle =\langle \left(-\mathbf 1_{\star}\right)(\mathcal{D}_1) \rangle =\left(-\mathbf 1_{\star}\right)(\langle \mathcal{D}_1\rangle)=\left(-\mathbf 1_{\star}\right)(\mathcal{P}_1)$. The   equality $\widehat{\mathcal{Q}_i}^{\perp}=-\mathbf 1_{\star}(\widehat{\mathcal{P}_i}) $ can be proven in a similar way.  For proving  the second statement  in  the theorem we consider Theorem ~\ref{forproof} item 6) and the equalities there for $\widehat{\mathcal{P}_i}$ and $\widehat{\mathcal{Q}_i}$. if $-\mathbf 1_{\star}(E_i)=E_i$ then $\mathcal{Q}_i\cap \mathcal{Q}^\perp_i=\mathcal{Q}_i\cap \mathcal{P}_i=\{0\}$ and $\widehat{\mathcal{Q}_i}\cap \widehat{\mathcal{Q}}^\perp_i=\widehat{\mathcal{Q}_i}\cap \widehat{\mathcal{P}_i}=\{0\}$. So $\mathcal{Q}_i$  and $\widehat{\mathcal{Q}_i}$ are LCD codes over $\mathcal{R}$, for $ 1\leq i \leq m$. Similarly, the case for $ \mathcal{P}_i,\widehat{\mathcal{P}_i}$  also holds.
    \end{proof}

%It is clear that if we have $E_1,\ldots, E_m$ idempotents in $R[A]$  defined as in Section~\ref{sec:spl} then 
%$$ \sum_{i=1}^{m} \theta_{A_i}E_i $$ 
%is an idempotent in $\mathcal R[A]=R[X_1,\ldots, X_s,Y_1,\ldots, Y_\delta]/\langle I  , I_A\rangle$.

\section{Polyadic consta-abelian codes over serial rings}\label{sec:core2}

In this section, we will take $\Theta=A$ and we will consider splitting as in Definition~\ref{def:classes}. Recall the set $\Bar{S}=\{1+r(s-1)|s \in S\}$ Section \ref{sec:polyconsta} and the associated cyclotomic class $C_{\Bar{S}}$. As in the previous section \ref{sec:core1} we have the partition of the class $C_{\Bar{S}}$   as  
\begin{equation} \label{eq:mpartition2}
     C_{\Bar{S}}= \{ C_i \mid i=1,\ldots, | C_{\Bar{S}}|\}= A_1\cup \dots\cup A_m
\end{equation}
We have two different cases depending on whether the $|C_S|$   is greater than $m$ or not. Let $|A_i|=t_i$, $i=1,\ldots, | C_{\Bar{S}}|$. If $|{C}_{\Bar{S}}|\geq m$  for $1\leq t_i \leq |{C}_{\Bar{S}}|-m+1$, each  $A_i$ is non-empty set. Otherwise, $|{C}_{\Bar{S}}|$ sets in the partition in the Equation~\ref{eq:mpartition2} are non-empty and they have only one element and the remaining $m-|{C}_{\Bar{S}}|$ are empty.  It can be easily checked
that $|{C}_{\Bar{S}}|=\sum_{i=1}^{m}t_i$. We will represent the idempotents of polyadic consta-Abelian codes defined on $ R[A,\bm\lambda]$ with $D_i$, $\widehat{D}_i$, $E_i$ and $\widehat{E}_i$. In the case of Type I, recall that there should be only one type idempotent and its pair, say $D_i$ and $\widehat{D}_i$. Now we can define the polyadic consta-Abelian codes over $\mathcal R$ using p of polyadic consta-Abelian codes over a chain ring $R$  Thus we get the following codes  (in this case, we have also Type I codes since $S_\infty$ can be the empty set) over  $\mathcal R[A,\bm\lambda]=R[X_1,\ldots, X_s,Y_1,\ldots, Y_\delta]/\langle I  , I_{A,\bm\lambda}\rangle$. 

\begin{definition}\label{def:types} Let $\Theta=A$  and an  $\mathcal S=(S_\infty, S_0, S_1, ..., S_{m-1})$ an  $m-$ splitting of $\Theta$ w.r.t. $r$. Let $k_{ij}$ be integers such that $k_{ij}=i-j+1 $ $mod \  (m)$. Let $L_i$, $\widehat{L_i}$, $K_i$ and  $\widehat{K_i}$ be defined as in Section~\ref{sec:polyconsta}.
     \begin{enumerate}\item \underline{Type I codes}

The idempotent generators of the polyadic codes over the ring $\mathcal R[A,\bm\lambda]$ for each $j=2,\ldots, m$
 
     \begin{itemize}
        \item $\mathcal{D}_1=\sum_{i=1}^{m} \theta_{A_i}D_i $ where $D_i$ is the idempotent generator for $L_i$.
        \item $\mathcal{D}_j=u_*(\mathcal{D}_{j-1})=\sum_{i=1}^{m} \theta_{A_i}D_{k_{ij}} $
           \end{itemize}

Let the polyadic consta-abelian codes associated with the idempotents $\mathcal{D}_j$ over $\mathcal R[A,\bm\lambda]$ be called as $\mathcal{P}_j$. So, the desired polyadic consta-abelian code is generated by the idempotent such that $\mathcal{P}_j=\langle \mathcal{D}_j \rangle $. 
    
\item \underline{Type II codes}

\begin{itemize}

    \item Odd-like idempotent generators over the ring $\mathcal R[A,\bm\lambda]$ for each $j=2,\ldots, m$
    \begin{itemize}
        \item $\mathcal{D}_1=\sum_{i=1}^{m} \theta_{A_i}D_i $ where $D_i$ is the idempotent generator for $L_i$.
        \item $\mathcal{D}_j=u_*(\mathcal{D}_{j-1})=\sum_{i=1}^{m} \theta_{A_i}D_{k_{ij}} $
        \item $\mathcal{D}_1'=\sum_{i=1}^{m} \theta_{A_i}D_i' $ where $D_i^\prime$ is the idempotent generator for $\widehat{L_i}$.
        \item $\mathcal{D}_j'=u_*(\mathcal{D}_{j-1}')=\sum_{i=1}^{m} \theta_{A_i}D'_{k_{ij}} $ 
    \end{itemize}
    \item Even-like idempotent generators over the ring $\mathcal R[A,\bm\lambda]$ for each $j=2,\ldots, m$
    \begin{itemize}
        \item  $\mathcal{E}_1=\sum_{i=1}^{m} \theta_{A_i}E_i $ where $E_i$ is the idempotent generator for $K_i$.
        \item $\mathcal{E}_j=u_*(\mathcal{E}_{j-1})=\sum_{i=1}^{m} \theta_{A_i}E_{k_{ij}} $ 
        \item $\mathcal{E}_1'=\sum_{i=1}^{m} \theta_{A_i}E_i' $ where $E_i$ is the idempotent generator for $\widehat{K_i}$.
        \item $\mathcal{E}_j'=u_*(\mathcal{E}_{j-1}')=\sum_{i=1}^{m} \theta_{A_i}E'_{k_{ij}} $ 
    \end{itemize}
\end{itemize}

Let the polyadic consta-abelian codes associated with the idempotents $\mathcal{D}_j$, $\mathcal{D}_j'$, $\mathcal{E}_j$ and $\mathcal{E}_j'$ over $\mathcal R[A,\bm\lambda]$ be called as $\mathcal{P}_j$, $\widehat{\mathcal{P}_j}$, $\mathcal{Q}_j$ and $\widehat{\mathcal{Q}_j}$, respectively. So, the desired polyadic consta-abelian codes are generated by the idempotents such that $\mathcal{P}_j=\langle \mathcal{D}_j \rangle $, $\widehat{\mathcal{P}_j}=\langle \mathcal{D}_j' \rangle $, $\mathcal{Q}_j=\langle \mathcal{E}_j \rangle $ and $\widehat{\mathcal{Q
 }_j}=\langle \mathcal{E}_j' \rangle $. 
\end{enumerate}
\end{definition}

It is a straightforward exercise to check that the following results can be proven in the same fashion as the Abelian case in Section~\ref{sec:core1} above since they only rely on the decomposition of the idempotents in the polynomial rings $R[X_1,\ldots, X_s]/\langle I\rangle$ and $R[Y_1,\ldots, Y_\delta]/\langle   I_{A,\bm\lambda}\rangle$.  The first two theorems are related to type I codes and the remaining ones to type II codes.

\begin{theorem} [Number of polyadic consta abelian codes of Type I]  Consider Type I codes in Definition~\ref{def:types}. The following statements hold.

\begin{enumerate}
    \item If  $|C_{\Bar{S}}| \geq m $, then the number of inequivalent polyadic consta Abelian codes of Type I over the ring $\mathcal{R}{[A,\bm\lambda]}$ is equal to {  $$ \frac{1}{m} \sum_{t_{m-1}=1}^{|C_{\Bar{S}}|-T_{m-2}-1} \ldots \sum_{t_{2}=1}^{|C_{\Bar{S}}|-T_1-(m-2)}\sum_{t_{1}=1}^{|C_{\Bar{S}}|-(m-1)} \binom{|C_{\Bar{S}}|}{t_1} \binom{|C_{\Bar{S}}|-t_1}{t_2} \ldots \binom{|C_{\Bar{S}}|-(T_{m-2})}{t_{m-1}}   $$} where $T_i=\sum_{j=1}^{i} t_j$.
    
    \item If $|C_{\Bar{S}}| < m $, then the number of inequivalent polyadic consta-abelian codes of Type I over the ring $\mathcal{R}{[A,\bm\lambda]}$ is equal to
    $$ \frac{1}{m} (|C_{\Bar{S}}|)!\binom{m}{|C_{\Bar{S}}|}  $$
    
\end{enumerate}
\end{theorem}
\begin{proof}
    Considering the same counting argument as in  Theorem~\ref{theo:number}, it can  be seen that the number of the polyadic consta-Abelian codes of Type I is equal to  half of the number for old ones since there is only one type idempotent. 
\end{proof}

\begin{theorem} \label{forproof2}
    Let $B$ be a subset of $\{1,2,\ldots,m\}$  with at least two elements.  The following propositions are satisfied for polyadic consta-Abelian codes of Type I $\mathcal{P}_i$ over $\mathcal{R}{[A,\bm\lambda]}$ defined as above.  
    \begin{enumerate}

          \item $\bigcap_{i=1}^m\mathcal{P}_i=\bigcap_{j\in B}\mathcal{P}_j=\{0\}$
        \item $\sum_{i=1}^m\mathcal{P}_i=\mathcal{R}[A,\bm\lambda]$
        \item $\prod_{i=1}^m {\mathcal{D}_i}=\prod_{j\in B} {\mathcal{D}_j}=\{0\}$
        \item $\sum_{i=1}^m {\mathcal{D}_i}=1$
    \end{enumerate}
    
\end{theorem}

\begin{theorem} [Number of polyadic consta-abelian codes of Type II] \label{theo:number3} Consider Type I codes in Definition~\ref{def:types}. The following statements hold.

\begin{enumerate}
    \item If  $|C_{\Bar{S}}| \geq m $, then the number of inequivalent odd-like (or even-like) polyadic codes over the ring $\mathcal{R}{[A,\bm\lambda]}$ is equal to {  $$ \frac{2}{m} \sum_{t_{m-1}=1}^{|C_{\Bar{S}}|-T_{m-2}-1} \ldots \sum_{t_{2}=1}^{|C_{\Bar{S}}|-T_1-(m-2)}\sum_{t_{1}=1}^{|C_{\Bar{S}}|-(m-1)} \binom{|C_{\Bar{S}}|}{t_1} \binom{|C_{\Bar{S}}|-T_1}{t_2} \ldots \binom{|C_{\Bar{S}}|-T_{m-2}}{t_{m-1}},   $$} where $T_i=\sum_{j=1}^{i} t_j$.
    
    \item If $|C_{\Bar{S}}| < m $, then the number of inequivalent odd-like (or even-like) polyadic codes over the ring $\mathcal{R}{[A,\bm\lambda]}$ is equal to
    $$ \frac{2}{m} (|C_{\Bar{S}}|)!\binom{m}{|C_{\Bar{S}}|}.  $$
    
\end{enumerate}

\end{theorem}

\begin{theorem} \label{forproof3}
    Let $B$ be a subset of $\{1,2,\ldots,m\}$ with  at least two elements.  The following propositions are satisfied for polyadic consta-Abelian codes of Type II $\mathcal{P}_i$ and $\mathcal{Q}_i$ over $\mathcal{R}{[A,\bm\lambda]}$  given in   Definition~\ref{def:types}.  
    \begin{enumerate}
        \item $\bigcap_{i=1}^m\mathcal{P}_i=\mathrm{Rep}(n)$, the repetition code over $\mathcal{R}{[A,\bm\lambda]}$
        \item $\sum_{i=1}^m\mathcal{P}_i=\sum_{j\in B}\mathcal{P}_j=\mathcal{R}[A,\bm\lambda]$.
        \item $\bigcap_{i=1}^m\mathcal{Q}_i=\bigcap_{j\in B}\mathcal{Q}_j=\{0\}$.
        \item $\sum_{i=1}^m\mathcal{Q}_i=\mathrm{Rep}(n)^\perp$ .
        \item $\mathcal{Q}_i\cap \mathrm{Rep}(n) =\{0\}$ and $\mathcal{P}_i\cap \mathrm{Rep}(n) =\mathrm{Rep}(n)$ for $ 1\leq i \leq m$.
        \item $\mathcal{P}_i+\mathcal{Q}_i=\mathcal{R}[A,\bm\lambda]  $ and  $\mathcal{P}_i\cap \mathcal{Q}_i=\{0\}$ for $ 1\leq i \leq m$.
    \end{enumerate}
    If we consider $\widehat{\mathcal{P}_i}$'s  and $\widehat{\mathcal{Q}_i}$'s instead of  $\mathcal{P}_i$'s   and $\mathcal{Q}_i$'s respectively, the previous statements also hold.
\end{theorem}
\begin{proof}
    Just follow the  proof of Theorem~\ref{forproof} using consta-Abelian polyadic code definition.

   \end{proof}

\begin{theorem}
    Let $B$ be a subset of $\{1,2,\ldots,m\}$ with at least two elements.  The following statements are satisfied for polyadic codes $\mathcal{P}_i$ and $\mathcal{Q}_i$ over $\mathcal{R}$  given in   Definition~\ref{def:types}.   
    \begin{enumerate}
        \item $\mathcal{Q}_i+\mathrm{Rep}(n)=\widehat{\mathcal{P}_i}$ and $\widehat{\mathcal{Q}_i}+\mathrm{Rep}(n)=\mathcal{P}_i$ 
        \item $\mathcal{Q}_i \cap \widehat{\mathcal{Q}_i}=\mathrm{Rep}(n)^\perp$ and $\mathcal{Q}_i \cap \widehat{\mathcal{Q}_i}=\{ 0 \}$
        \item  $\mathcal{P}_i+\widehat{\mathcal{P}_i}=\mathcal{R}[A,\bm\lambda] $ and $\mathcal{P}_i \cap \widehat{\mathcal{P}_i}=\mathrm{Rep}(n)$

    \end{enumerate}
\end{theorem}
In the nega-Abelian case, that is if $\bm \lambda=(-1,\ldots, -1)$, a result like Theorem \ref{the:LCD1} in the previous section can be proven.

 \begin{theorem} [Nega-Abelian LCD codes]
    Let $\mathcal{Q}_i $ and $\widehat{\mathcal{Q}_i} $ be a pair of even-like polyadic negacyclic codes of Type II with the associated odd-like polyadic negacyclic codes of Type II $\mathcal{P}_i $ and $\widehat{\mathcal{P}_i} $ over the ring $\mathcal{R}{[A,\bm\lambda]}$ for    $ 1\leq i \leq m$. The following statements are hold: 
    \begin{enumerate}
        \item $\mathcal{Q}_i^{\perp}=-\mathbf 1_{\star}(\mathcal{P}_i) $ and $\widehat{\mathcal{Q}_i}^{\perp}=-\mathbf 1_{\star}(\widehat{\mathcal{P}_i}) $
        \item If $-\mathbf 1_{\star}(E_i)=E_i$  then $\mathcal{Q}_i^{\perp}=\mathcal{P}_i$, $\widehat{\mathcal{Q}_i}^{\perp}=\widehat{\mathcal{P}_i}$ and $\mathcal{Q}_i, \widehat{\mathcal{Q}_i}, \mathcal{P}_i,\widehat{\mathcal{P}_i}$ are LCD codes over $\mathcal{R}{[A,\bm\lambda]}$, for $ 1\leq i \leq m$.
    \end{enumerate}

\end{theorem}

\section{Conclusions}\label{sec:conclude} In this paper we have studied  polyadic  Abelian codes and consta-Abelian codes  defined over  some serial rings defined as affine algebras of a certain type  with a finite commutative chain
              coefficient ring.   We have completely described them in terms of their generators associated with  the concrete splitting of the Abelian group underlying the structure. As a follow-up applied work,  it will be nice to 
 check if the  Gray mappings associated with the idempotent decomposition of the codes (see \cite{Indian,Indian2,Habibul}) provide  codes with good properties over the base chain ring.
\bibliographystyle{plain}
\bibliography{refs.bib}
	
\end{document}